# Signal to Noise and b-value Analysis for Optimal Intra-Voxel Incoherent Motion Imaging in the Brain

Harri Merisaari[1,2], Christian Federau[3]

[1] Department of Radiology, University of Turku, Turku, Finland

[2] Department of Future Technologies, University of Turku, Finland

[3] Institute for Biomedical Engineering, ETH, Zürich and University Zürich, Zürich, Switzerland

Corresponding author: Harri Merisaari
E-mail address: harri.merisaari@utu.fi
Full Postal address:
Harri Merisaari
Department of Radiology
Turku University
Kiinamyllynkatu 4-8
20520, Turku, Finland

# ABSTRACT

Intravoxel incoherent motion (IVIM) is a method that can provide quantitative information about perfusion in the human body, in vivo, and without contrast agent. Unfortunately, the IVIM perfusion parameter maps are known to be relatively noisy in the brain, in particular for the pseudo-diffusion coefficient, which might hinder its potential broader use in clinical applications. Therefore, we studied the conditions to produce optimal IVIM perfusion images in the brain. IVIM imaging was performed on a 3-Tesla clinical system in four healthy volunteers, with 16 b values 0, 10, 20, 40, 80, 110, 140, 170, 200, 300, 400, 500, 600, 700, 800, 900 s/mm$^2$, repeated 20 times. We analyzed the noise characteristics of the trace images as a function of b-value, and the homogeneity of the IVIM parameter maps across number of averages and sub-sets of the acquired b values. We found two peaks of noise of the trace images as function of b value, one due to thermal noise at high b-value, and one due to physiological noise at low b-value. The selection of b value distribution was found to have higher impact on the homogeneity of the IVIM parameter maps than the number of averages. Based on evaluations, we suggest an optimal b value acquisition scheme for a 12 min scan as 0 (7), 20 (4), 140 (19), 300 (9), 500 (19), 700 (1), 800 (4), 900 (1) s/mm$^2$.

# 1 INTRODUCTION

Intravoxel incoherent motion (IVIM) is a method to separate perfusion effects from thermal diffusion effects from images acquired using diffusion-weighted magnetic resonance (Le Bihan et al., 1988). A relatively large amount of experimental evidence consistent with the interpretation that the IVIM method can provide in vivo perfusion information has been collected in the last few years (Christian Federau, 2017). In particular, the IVIM method has been shown to be applicable in a broad range of brain clinical investigations (Christian Federau, O'Brien, et al., 2014), both in the context of hyperperfused lesions such as in high-grade glioma (Bisdas, 2013; Catanese et al., 2018; Christian Federau, Meuli, et al., 2014; Gielen et al., 2017; Keil et al., 2015; Shen et al., 2016; Togao et al., 2016; Wang et al., 2019; Zou et al., 2018), and hypoperfused lesions such as strokes (C. Federau et al., 2014; Christian Federau et al., 2019; Suo et al., 2016; Yao et al., 2016), vasospasm (Heit et al., 2018), cerebral lymphoma (Yamashita et al., 2016) and cerebral death (Christian Federau et al., 2016). In addition, the method has shown promises for the survival prognosis in high-grade brain glioma (Christian Federau et al., 2017; Puig et al., 2016), in differentiating recurrent tumour from radiation necrosis for brain metastases treated with radiosurgery (Detsky et al., 2017), and as a surrogate marker for the progression of cerebral small vessel disease (Sau May Wong et al., 2017; Zhang et al., 2017). Unfortunately, IVIM perfusion parameters maps are known to be noisy (Ahlgren et al., 2016; Merisaari et al., 2017; Wu et al., 2015), and this is particularly harmful for the detection of hypoperfused lesions, because the quality of the IVIM signal equation fit decreases with decreasing perfusion fraction. Optimizing the acquisition parameters might help reduce this drawback.

Several studies on the effect of b-value distribution on the IVIM reconstruction have been conducted. However, to the best of our knowledge no exhaustive evaluation for the optimal choice of b value in the brain has been conducted, considering the number of averaged repetitions. Lemke et al. studied b value distribution in liver (Lemke et al., 2011). In the brain, Chabert et al. studied 10 subjects with two b value distributions (Chabert et al., 2019). Hu et. al (Hu et al., 2020) studied retrospectively 22 healthy males with 12 b-value sets in low and high groups, suggesting total of eight b-values up to 800-1000 s/mm$^2$. A further study in eight healthy subjects with an acquisition with four repetitions in the upper abdomen, an optimization derived from based on Cramér-Rao lower bounds suggested to use twice as many b-values as b=0 images (Birkbeck et al., 2012; Qinwei Zhang et al., 2013). In (Meeus et al., 2018), simulations and 16 healthy volunteers were analyzed in a rapid measurement setting evaluating two b value distributions. Reproducibility across sites and scanner models with selected b-value set for two of IVIM parameters *f* and *D* parameters only, was evaluated in (Grech-Sollars et al., 2015). Further, inter-reader reliability with eight subjects in various organs, including brain, was studied in (Filli et al., 2015), and short-term repeatability in (Stieb et al., 2016).

The purpose of this study was to quantify in the brain the conditions to produce optimal IVIM perfusion images. For this, we acquired 20 averages of 16 b values ranging from 0 to 900 s/mm$^2$ during a 1-hour scan in four volunteers. We studied the noise characteristics of the trace images in a large number of sub-sets of the acquired b values and number of averages. Finally, we studied and optimized the signal-to-noise properties of the IVIM perfusion maps, using various b set selection strategies.

# 2 MATERIALS AND METHODS

*2.1 Data acquisition*

IVIM imaging was performed on a 3-Tesla clinical system (Siemens, Erlangen, Germany) in four healthy volunteers (1 female 25 y, 3 males 26, 38, and 38 y) with parameters: TR/TE 4000/92 ms,

FOV 22x22 cm, matrix size 148x148, for a voxel size of 1.5x1.5x6 mm$^3$ in approval of local ethics committee. The diffusion weighting was applied in three orthogonal directions with 16 b values 0, 10, 20, 40, 80, 110, 140, 170, 200, 300, 400, 500, 600, 700, 800, 900 s/mm$^2$, from which the trace was built. The acquisition was repeated 20 times and images with number of averages 1-5 of the measurements were calculated so that in the analysis there was 20 images including 1 repetition, 10 average images each containing 2 repetitions, and correspondingly for average image containing 3,4 and 5 repetitions each. The complete analysis flow is shown in **Figure 1**. All patients included in the study provided informed consent, and the study was permitted by the local ethics committee.

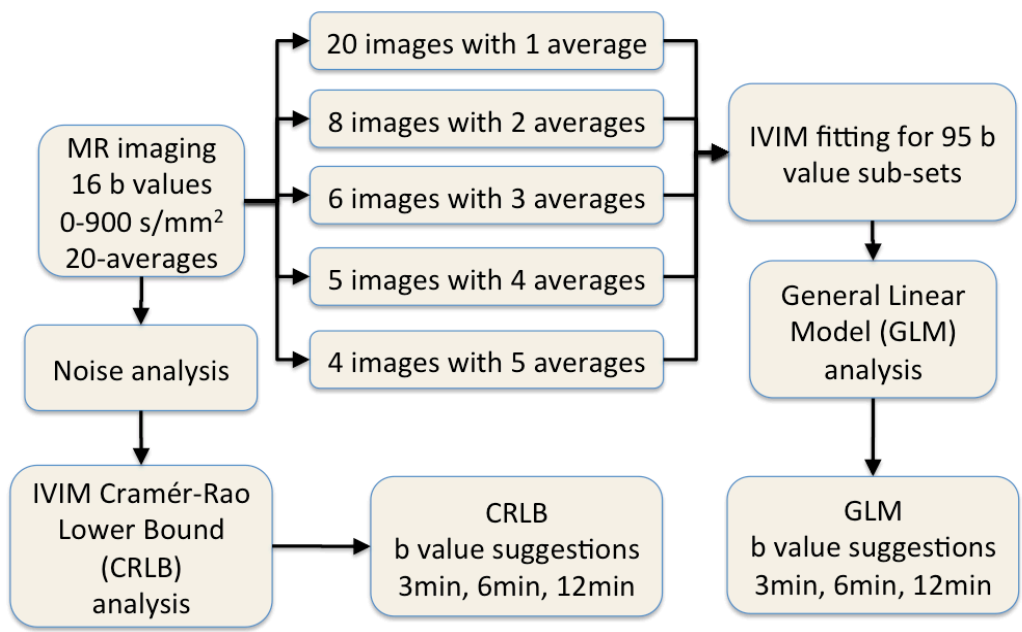

**Figure 1** Analysis workflow for optimization of brain IVIM model for number of b values up to 900 s/mm$^2$, using 20 repeated scans in four healthy volunteers, with two optimization strategies of General Linear Modelling (GLM) of standard deviation measurements in 95 b value settings, and Cramér-Rao Lower Bound analysis using noise measurements in the same data.

*2.2 IVIM model fitting*

The IVIM imaging data were post-processed using FSL (FMRIB's Software Library, www.fmrib.ox.ac.uk/fsl), and locally written python and C++ code. The b0 images were co-registered to the first b0 image. The co-registered b0 image were averaged, after which all images were co-registered to the averaged b0 to assess motion during scanning session. The DWI decay curves were fitted with Broyden–Fletcher–Goldfarb–Shanno (BFGS) algorithm implemented in the Dlib-ml C++ library (18). We followed the fitting procedure described in (Merisaari et al., 2017), the fittings were initialized with slopes and perfusion fraction values calculated analytically from the log-transform of the signal and an extrapolated line at b=0, from the tail values selected according to the available b values (see below) in the images so that all b values >= 300 s/mm$^2$ were included. The *D* parameter was fitted first with the values at the tail of the decay curve, and then fixed for the fitting of the *f* and *D\** parameters. This provided two-stage fittings of the curve with two and three degrees of freedom in the two stages of the fitting procedure, correspondingly. It is to be noted, that in special case of evaluating only three b-values, the fitting procedure is reduced to be identical to determination of initialization of values, as there is no other data point available for the fitting.

*2.3 Analysis Set-up*

Post-processing was done for all acquired images 1-20, and for a subset of all possible sets of b values arising from the acquired 16 b values (to limit the number of evaluations and generated parameters maps with this data to a reasonable amount). A large ensemble of different possible combinations of b values was produced as follow: From the initial set of 16 b values acquired, 3 strategies of variable b values were selected, starting from the same set of 3 b values: b=0, 200, and 900 s/mm$^2$.

- In the first set, b values below 200 s/mm² were selected, following the strategy of maximum sampling in a b value range of the IVIM effect.
- In the second strategy, b values above 200 s/mm² were selected to follow the approach of maximizing good starting estimate of *D* (Rydhög et al., 2014).
- Finally, the third strategy alternated one value above 200 s/mm² and one value below 200 s/mm², to give emphasis in the mid-range of b-values, as study in (Merisaari & Jambor, 2015), has been shown to improve accuracy in estimation of fraction *f* between two exponential decays.

The 3 strategies of b values were then pooled in one ensemble, and further augmented with manually selected b value sets to homogenize the sets and to evaluate previous work on IVIM brain (Meeus et al., 2018), to a final ensemble of 95 different sets of b values (**Supplementary Material Appendix A**). The selected b values thus roughly represent three b value selection strategies of preferring low b values instead of high ones, preferring high b values instead of low ones, and balancing added b values between low and high b values, and correspond in large part of the b value settings proposed in the literature, within the limits of possible b value sub-sets which could be generated from the acquired b values. Low b values were defined as the values below 200 s/mm². Values 200, 900 were fixed to follow suggestions of earlier b value set optimization studies (Chabert et al., 2019; Meeus et al., 2018), and were considered as suggested optimal for optimizing *f* parameter.

*2.4 Brain segmentation*

The brain was segmented in grey matter (GM) and white matter (WM) with averages of b0 image in 1-hour acquisition, by thresholding followed by manual edits to remove artifacts. The voxels of the GM and the WM maps were then pooled for further analysis. The WM region was eroded with one voxel before extracting voxel values, to address potential partial volume effect from neighbouring regions.

*2.5 Standard deviation analysis inside the GM and WM* of the original data

To get an overview of the noise characteristics of the original data, the coefficient of variance (i.e. standard deviation divided by the mean) of the voxel intensity values in the GM and WM was calculated for each b value and each average (i.e. 1 to 20 averages) trace images, and plotted as heatmaps for the first of the scanned subjects.

*2.6 Standard deviation analysis inside the GM and WM* of the IVIM parameters

The standard deviation of the IVIM parameters (*f*, *D*, *D\**) in GM and WM were calculated for the sets of b values and the averaged images containing 1-5 repetitions. Each set of different b value was analyzed separately for the number of repetitions in average image ranging from 1 to 5. The number of average images were: 20 for 1-average, 10 for 2-average, 6 for 3-average, 5 for 4-average, and 4 for 5-average, total 45 images for each b value setting. The corresponding numbers of b0 images were 1, 2, 3, 4 and 5. The b0 images were averaged in the latter four partitions.

*2.7 Optimal b value set and the number of averages for a given scan time of 3 min, 6 min and 12 min.*

Finally, we assessed the optimal set of b values and number of averages to produce the most homogenous *D\** maps possible, from all subsets analyzed (see above and supplementary Material **Appendix A**). Under the constraint of a scan time below 3 min, 6 min and 12 min, we conducted regression analysis (General Linear Model, GLM) to see effect of individual b-values and b-value sets to standard deviation of parameter maps in GM and WM. In addition to regression analysis with four subject and 95 different b-value set, we performed Cramér-Rao Lower Bound (CRBM) analysis (Su & Rutt, 2014) with using measured noise in b-values range 0-900 s/mm$^2$ across repetitions, optimizing for all IVIM parameters together and allowing individual b values to have different number of averages to each other.

*2.8 Statistical analysis*

Statistical significance of monotonic upward or downward trends in number of averaged images and number of b values were tested with the Mann-Kendall trend test. Effect of b-value selection, number of repetitions, and the number of b-values were used in multivariate analysis to analyze the relative contribution of these variables to standard deviation IVIM parameter maps. P-values less than 0.05 after Bonferroni correction were considered statistically significant while raw p-values are reported unless otherwise noted. All statistical tests were done in RStudio environment (v 1.1.383, 2017 RStudio, Inc.).

# 3 RESULTS

*3.1 Standard deviation inside GM and WM*

The coefficient of variance (CoV) as a function of b value and number of average in the GM and WM (**Figure 2**) varied in 15%-40% for GM and 9%-10.5% for WM. WM CoV had two peaks at the low number of b-values: One at high b values, due to thermal noise, and interestingly, another one at low b values, most probably due to physiological noise arising from variation in blood flow during the cardiac and respiratory cycles. Interestingly, but not surprisingly, the latter peak was more prominent in the GM compared to the WM, which might probably be due to the known fact that the capillary network is more prominent in the GM. We observed that this increase in SD is due to the inclusion of low b-value < 200 s/mm$^2$ with a larger signal SD due to physiological noise (see **Figure 2**), and that number of repetitions did not bring notable improvement to it in four evaluated cases and repetitions up to 20.

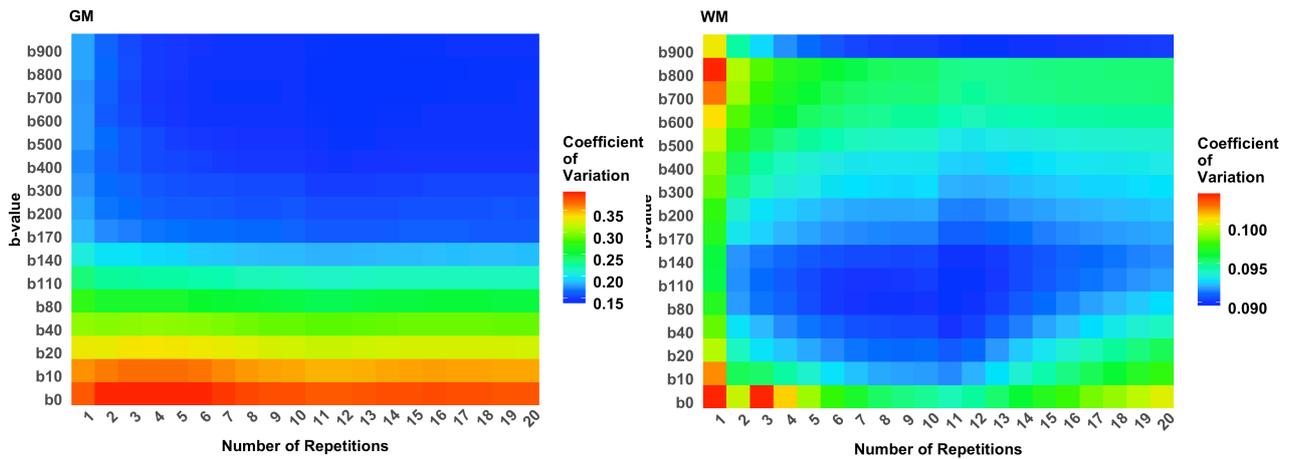

**Figure 2** Coefficient of Variation (SD/mean) as a function of b values and number of averages, in the gray matter (left) and the white matter (right) for four healthy volunteers in DWI trace images in b-values ranging from 0 to 900 s/mm$^2$ and number of repetitions for signal averaging in 1 to 20. Gray matter variation begins to drop after b=170 s/mm$^2$ and with 4 repetitions, while white matter variation is generally less and with negligible differences between b-values and repetitions in range 9% to 10.5%, and lowest points around b=80-140 s/mm$^2$ and when using 4 to 13 averages.

*3.2 Effect of number of averages and effect of b-value selection in four subjects*

When b-value sets having the same number of b-values were averaged and four subjects were combined, interaction between the number of b-values and number of repetitions in WM $D$ ($p<1.0\times10^{-6}$), and effect of adding variation between subjects to the model in $D^*$ and $f$ ($p<1.0\times10^{-6}$) were found to be significant and were addressed in the subsequent analysis. We observed that number of b-values was affecting SD of WM and GM in IVIM parameter maps than averaging of the signal according to trend test over all four subjects ($p<1.0\times10^{-6}$). Monotonic trend was statistically significant in WM and GM for all three variables for the number of b-values ($p<1.0\times10^{-6}$) and for the number of repetition ($p<0.007$). Overall with all b-value sets and number of averages, when testing the effect of an individual b-value set to SD, selection of individual b-value set was explaining SD significantly better than signal averaging ($p<1.0\times10^{-10}$).

We evaluated the b value sets and b values individually to see their effect on SD. The suggested b values which had statistical significance (p<0.001 after Bonferroni correction over b values) in explaining SD, and which made a statistically significant decreasing difference to SD are listed in **Table 1**. Generally, when other than the fixed b values of 0, 200, 900 s/mm² were considered in all b-value sets, the $D^*$ modelling was increasing SD dominantly with very small b values (10, 20, 40, 110, 600 s/mm²) in accordance with noise measurements in **Figure 2**. Also, high b-values (with one exception of 10, 110, 500, 600, 700 s/mm²) other than 900, increased SD of $f$. The same effect was visible when b values sets with a maximum scan time of 3, 6 and 12 minutes were considered. In CRLB analysis where optimized b value settings were aiming for all three IVIM parameters together, the suggested b values were largely in agreement from suggestion from direct SD evaluations with the ensemble of b value sets in GM and in WM. However, the CRLB analysis suggested different number of averages for selected b values, with non-uniform b value averaging and more emphasis towards using small b values with some of the highest available b values.

**Table 1** Suggested optimized b value sets for IVIM in the brain (number of averages in parenthesis), for 3 minutes, 6 minutes and 12 minutes scans for $f$, $D^*$, and $D$ parameters using fixed values for $f$, general linear model analysis for $D^*$ and $D$, and for all parameters together with Cramér-Rao Lower Bound estimation.

|  | 3 min scan | 6 min scan | 12 min scan |
|---|---|---|---|
| **Gray Matter** | | | |
| *f*-optimized | 0 (5), 200 (5), 900 (5) s/mm² | 0 (10), 200 (10), 900 (10) s/mm² | 0 (20), 200 (20), 900 (20) s/mm² |
| *D**-optimized | 0 (4), 200 (4), 800 (4), 900 (4) s/mm² | 0 (8), 200 (8), 800 (8), 900 (8) s/mm² | 0 (16), 200 (16), 800 (16), 900 (16) s/mm² |
| *D*-optimized | 0 (2), 10 (2), 20 (2), 40 (2), 80 (2), 140 (2), 200 (2), 900 (2) s/mm² | 0 (4), 10 (4), 20 (4), 40 (4), 80 (4), 140 (4), 200 (4), 900 (4) s/mm² | 0 (8), 10 (8), 20 (8), 40 (8), 80 (8), 140 (8), 200 (8), 900 (8)s/mm² |
| CRLB *f,D\*,D* | 0 (2), 20 (3), 110(2), 140 (2), 400 (1), 500 (4), 600 (2) s/mm$^2$ | 0 (2), 20 (5) 140 (6), 400 (9), 500 (1), 600 (5), 700 (1), 800 (1), 900 (2) s/mm$^2$ | 0 (7), 20 (4), 140 (19), 300 (9), 500 (19), 700 (1), 800 (4), 900 (1) s/mm$^2$ |
| **White Matter** | | | |
| *f*-optimized | 0 (5), 200 (5), 900 (5) s/mm$^2$ | 0 (10), 200 (10), 900 (10) s/mm$^2$ | 0 (20), 200 (20), 900 (20) s/mm$^2$ |
| *D\**-optimized | 0 (4), 200 (4), 800 (4), 900 (4) s/mm$^2$ | 0 (8), 200 (8), 800 (8), 900 (8) s/mm$^2$ | 0 (16), 200 (16), 800 (16), 900 (16) s/mm$^2$ |
| *D*-optimized | 0 (2), 10 (2), 20 (2), 40 (2), 80 (2), 200 (2), 300 (2), 900 (2) s/mm$^2$ | 0 (4), 10 (4), 20 (4), 80 (4), 200 (4), 300 (4), 900 (4) s/mm$^2$ | 0 (8), 10 (8), 20 (8), 40 (8), 80 (8), 200 (8), 300 (8), 900 (8) s/mm$^2$ |
| CRLB *f,D\*,D* | 0 (4), 20 (4), 110 (2), 140 (2), 400 (1), 600 (2), 700 (1) s/mm$^2$ | 0 (5), 20 (9), 40 (1), 80 (1), 110 (9), 140 (1), 170 (2), 500 (3), 600 (1) s/mm$^2$ | 0 (2), 40 (8), 140 (13), 170 (6), 300 (1), 400 (1), 500 (10), 600 (20), 800 (3) s/mm$^2$ |

*3.3 Optimal b value set and the number of averages for a given scan time of 3, 6 and 12 min.*

All optimized b value sets in **Table 1** were analyzed individually for their SD in GM and WM (*D\** **Figure 3, Supplementary Figures S1** and **S2** for *f* and *D*), and mean *f* (**Supplementary Figure S3**). We found negligible improvement between 3 min to 12 min for b value sub-set of 0, 200, 900 s/mm$^2$ which reflects direct analytical estimation of the IVIM parameters from the trace images without fitting, was providing reasonably good *D\** parameter maps in terms of SD, with some expence in quality of *f* values, while noting that low SD for those images may come in expense of losing true *D\** signal. Other b value sets which involve fitting, low SD in *D\** together with reasonable *f* was found with 12 min scan (0 (8), 10 (8), 20 (8), 40 (8), 80 (8), 140 (8), 200 (8), 900 (8) s/mm$^2$), and with two b value sub-sets suggested by CRLB for 6 min (0 (5), 20 (9), 40 (1), 80 (1), 110 (9), 140 (1), 170 (2), 500 (3), 600 (1) s/mm$^2$), and 12 min (0 (7), 20 (4), 140 (19), 300 (9), 500 (19), 700 (1), 800 (4), 900 (1) s/mm$^2$).

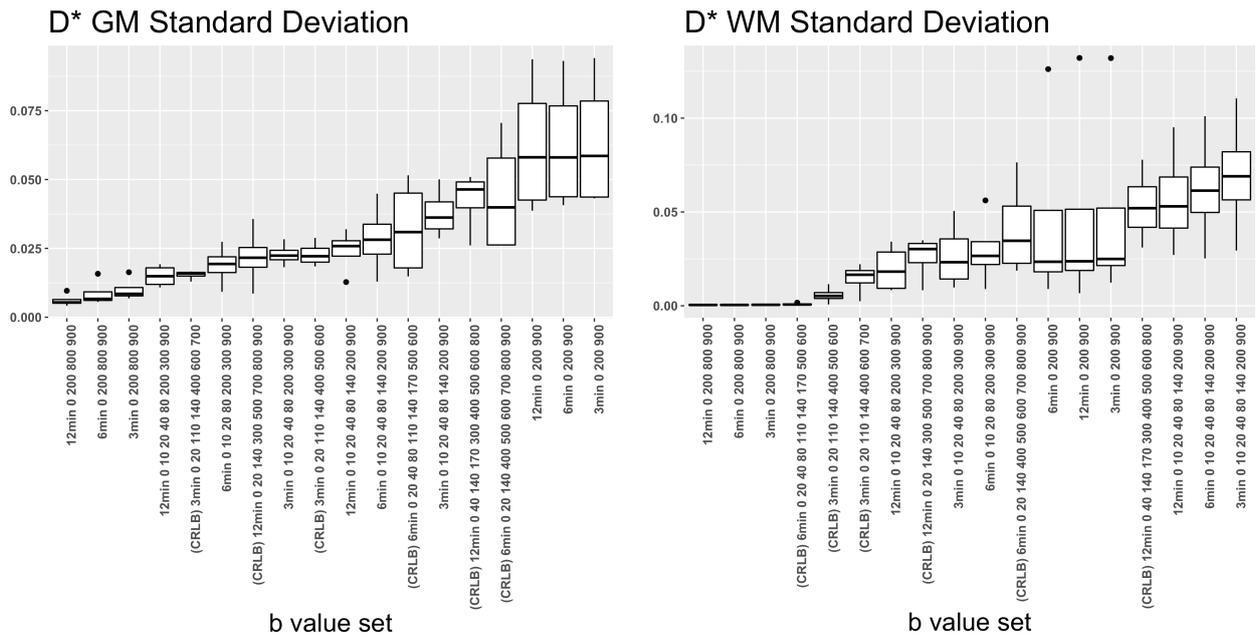

**Figure 3**: Standard deviations (SD) of IVIM parameter maps $D^*$ and $f$ in brain grey and white matter over four healthy volunteers, using optimized b value distributions for optimized b value settings from analysis of direct SD measurements, and CRLB using noise estimates, for 3, 6, and 12 minutes scans.

In comparison of parameter maps with the same b-value sub-sets, the 3 min, 6 min and 12 min $D^*$ and $f$ maps were similar to each other (**A-C** in **Figure 4** and **5**), with 0, 200, 900 s/mm$^2$, while there was apparent difference between optimized 12 min sub-set with GLM, and 6 min and 12 min sub-set with CRLB (**D-F** in Figure **4** and **5**). From all evaluated b-value subsets, the 12 min sub-set suggested from CRLB analysis provided most stable parameter maps, for both $f$ and $D^*$. The $D$ parameter maps quality followed the parameter map quality of $f$ (data not shown).

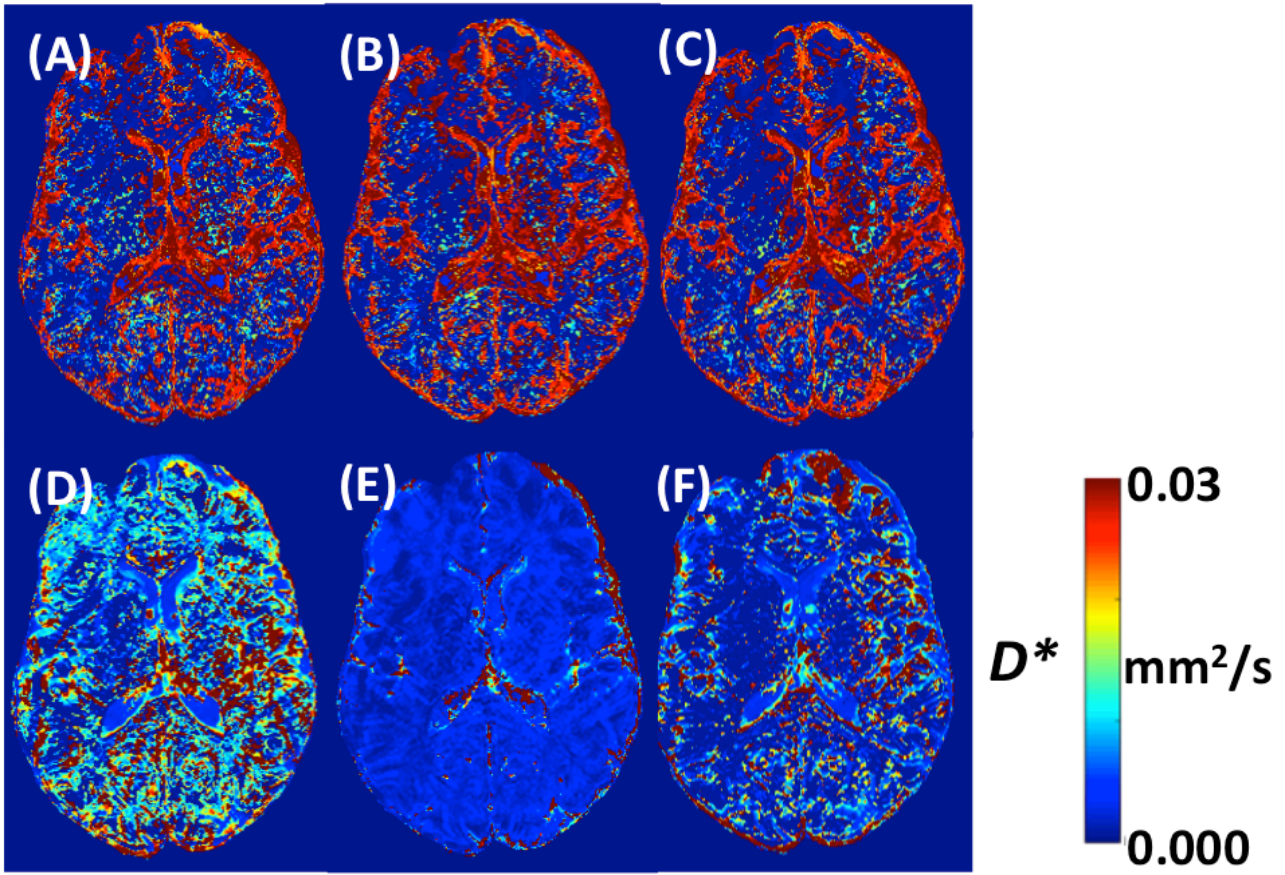

**Figure 4**: Example of IVIM $D^*$ parameter maps with acquisition schemes found best according to their standard deviation for gray and white matter. b values (number of averages) **(A)** 3 min (0 (5), 200 (5), 900 (5) s/mm$^2$), **(B)** 6 min (0 (10), 200 (10), 900 (10) s/mm$^2$), **(C)** 12 min (0 (20), 200 (20), 900 (20) s/mm$^2$), **(D)** 12 min (0 (8), 10 (8), 20 (8), 40 (8), 80 (8), 140 (8), 200 (8), 900 (8) s/mm$^2$), **(E)** 6 min (0 (5), 20 (9), 40 (1), 80 (1), 110 (9), 140 (1), 170 (2), 500 (3), 600 (1) s/mm$^2$), **(F)** 12 min (0 (7), 20 (4), 140 (19), 300 (9), 500 (19), 700 (1), 800 (4), 900 (1) s/mm$^2$).

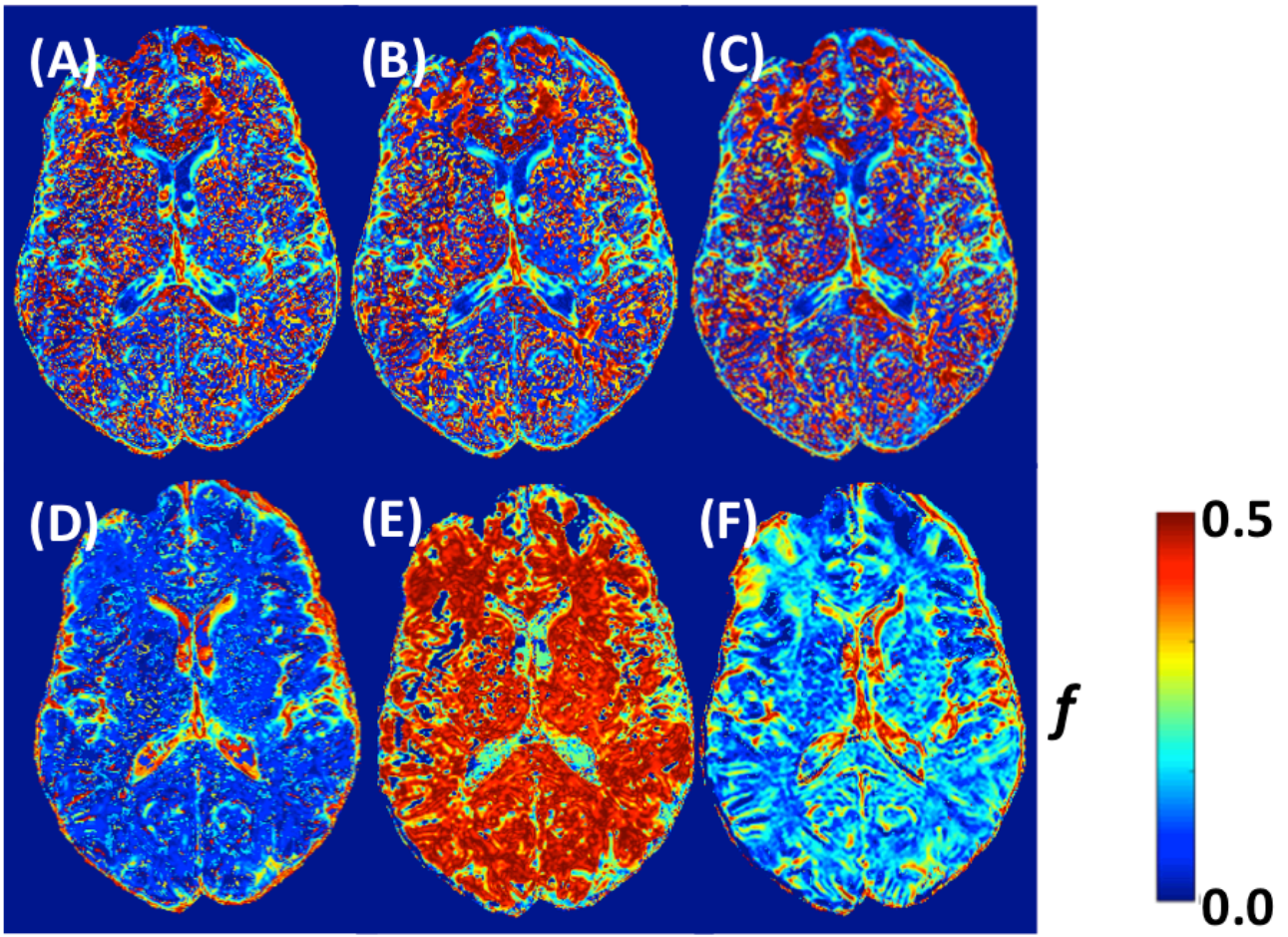

**Figure 5**: Example of IVIM *f* parameter maps with acquisition schemes found best according to their standard deviation for gray and white matter. b values (number of averages) **(A)** 3 min (0 (5), 200 (5), 900 (5) s/mm$^2$), **(B)** 6 min (0 (10), 200 (10), 900 (10) s/mm$^2$), **(C)** 12 min (0 (20), 200 (20), 900 (20) s/mm$^2$), **(D)** 12 min (0 (8), 10 (8), 20 (8), 40 (8), 80 (8), 140 (8), 200 (8), 900 (8) s/mm$^2$), **(E)** 6 min (0 (5), 20 (9), 40 (1), 80 (1), 110 (9), 140 (1), 170 (2), 500 (3), 600 (1) s/mm$^2$), **(F)** 12 min (0 (7), 20 (4), 140 (19), 300 (9), 500 (19), 700 (1), 800 (4), 900 (1) s/mm$^2$).

## 4 DISCUSSION

In this extensive study on the dependence of the homogeneity of IVIM parametric maps on the set of values and number of averages in the brain, we found an expected general trend toward an increase in image homogeneity with an increasing number of averages and a less trivial relationship between image homogeneity and the number of different b values used. We found that the inhomogeneity of the IVIM parametric maps increased significantly with the inclusion of b values smaller than 200 s/mm$^2$, which showed large SD due to physiological noise, probably mainly due to cardiac pulsation and to a lesser extent to the respiratory cycle. This effect was more pronounced in the GM compared to the WM, but was not observed for the parameter *D*, because the calculation of this parameter does not include values with physiological noise. *D* maps produced with two b values above 200 s/mm$^2$ and 2 averaged repetitions each showed already excellent and almost optimal image homogeneity, and only negligible improvements could be obtained if more repetitions or b values were added to the analysis, while *f* and *D\** benefitted from signal averaging especially when applied to lower b values < 200 s/mm$^2$. Overall, taking also into consideration the subjective aspect of the image, the set of parameters 12 min (0 (7), 20 (4), 140 (19), 300 (9), 500 (19), 700 (1), 800 (4), 900 (1) s/mm$^2$) seems to provide a good compromise to evaluate f and D* with reasonable low variation.

The choice of optimal scan parameters is in general not trivial, due to the interdependence of a relatively large number of parameters (such as scan time, resolution, TR, TE, bandwidth, for DWI the choice of the profile of the diffusion-sensitizing gradients) and effects (such as hardware related noise, eddy currents, field inhomogeneities, patient related physiological and motion artefacts). In addition, image homogeneity is not necessarily identical with holding maximal physiological or pathological information. Our analysis suggests that the set of b values should be selected with care for IVIM perfusion imaging, particularly when aiming for high-quality *D\** and *f* parameter maps, and that the optimal number of repetitions in average images differs between b values, due to need for addressing physiological noise in the low b values, and thermal noise in the high b values. In particular, it seems reasonable to suggest increasing the number of repetitions at low b values (<200s/mm$^2$) to average over physiological noise. In addition to compensation of physiological noise, placing more repetitions to mid-range of b values (200-400 s/mm$^2$) when making average images improves the estimation of fraction *f*, and *D\**. We speculate that mid-range b values are located where dominance of *D\** changes to *D*, and therefore they have an additional contribution in finding the fraction *f* between the two components of the model. A further option to consider to decrease physiological noise effects, although it increases scan time and is difficult to implement in the daily clinical routine because not very practical, is to use triggered acquisitions, such as cardiac gating (Christian Federau et al., 2013) and respiratory triggering. The use of triggered acquisition has already been shown to decrease measurement variability in IVIM liver imaging (Lee et al., 2015). Also, cerebrospinal fluid (CSF), which also undergoes periodic pulsations driven by cardiac and respiratory forces, and participate in the IVIM signal through partial volume (Becker et al., 2018; Surer et al., 2018), could be suppressed using an inversion-recovery pulse (Kwong et al., 1991; S. M. Wong et al., 2018), or even better, a T2-prepared inversion pulse (Christian Federau & O'Brien, 2015), which permits a better recovery of blood signal with similar suppression of the CSF signal.

There is to our knowledge no study on the choice of b value in the brain for IVIM perfusion imaging. Outside brain, typically 5 to 16 b values have been used to sample perfusion and diffusion IVIM effects (Taouli et al., 2016). Using Monte-Carlo simulation, Lemke et al. suggested an optimized set of 16 b values for the measurement of a low, medium and high perfusion fractions using Monte-Carlo simulations (Lemke et al., 2011). Cho et al. optimized the set b value using Monte-Carlo simulations and applied it successfully to IVIM parameters estimation in breast cancer (31). Pang et al. evaluated different combinations values in prostate cancer (Pang et al., 2013), Ter Voert et al in the liver (Ter Voert et al., 2016), and Dyvorne et al. found a subset of 4 optimized b values for the liver (Dyvorne et al., 2014).

Our study had several limitations. Only four subjects were scanned, while each case was analyzed extensively. Only one sequence of acquisition with maximum b value of 900 s/mm$^2$ was used, while sub-sets with the same acquired data were analyzed. Similarly, one acquired fitting approach was used, while the fitting procedure may affect the quality of measured IVIM parameter maps (Merisaari et al., 2017). In our fitting approach, the perfusion fraction and pseudo diffusion were not expected to affect the fit of $D$ (with >300 mm/s$^2$).

In conclusion, we evaluated the signal to noise dependence of IVIM trace image data and its effect on the quality of IVIM parameter maps of $D$, $f$ and $D^*$. We found that physiological noise at low b value and thermal noise at high b values propagates to the parameter maps. We suggest compensating for this effect by increasing the number of averages in those b values, with additional weighting at mid-range (200-400 s/mm$^2$) of b values. Overall, the set of parameters (0 (7), 20 (4), 140 (19), 300 (9), 500 (19), 700 (1), 800 (4), 900 (1) s/mm$^2$) seems to provide a good compromise to evaluate f and D* with reasonable low variation.

## ACKNOWLEDGEMENTS

Harri Merisaari was supported by the Cultural Foundation of Finland, and Orion Pharma Research Fellowship. Christian Federau was supported by the Swiss National Science Foundation (Grant No PZ00P3_173952 and CRSK-3_190697).